\documentclass[reprint, aps]{revtex4-2}

\usepackage{graphicx}
\usepackage{dcolumn}
\usepackage{bm,bbm}
\usepackage{amsmath,amssymb}
\usepackage[colorlinks,citecolor=blue]{hyperref}
\usepackage{ulem}

\begin{document}

\title{Aharonov-Bohm Effects for Electromagnetism and Gravity in Four-Dimensional Spacetime}

\author{Yanhui Li}
\email{Email: liyanhui0720@163.com} 
\author{Yakefu Reyimuaji}
\email{Email: yreyi@hotmail.com}
\affiliation{School of Physical Science and Technology, Xinjiang University,Urumqi 830017, China}

\date{\today}

\begin{abstract}
This paper investigates a geometric framework for the gravitational Aharonov-Bohm effect in four-dimensional spacetime, demonstrating how spacetime curvature induces nonlocal quantum phase shifts within field-free regions. By constructing vector bundles on spacetime manifolds equipped with Levi-Civita connections, we derive the holonomy transformations for parallel-transported quantum states. Under the Newtonian approximation, metric decomposition into Minkowski background plus scalar potential perturbations reveals through the linearized Einstein field equations that the gravitationally induced phase shift is mathematically isomorphic to its electromagnetic counterpart. These results establish the quantum observability of gravitational gauge structures and provide theoretical support for experimental verification via atom interferometry.

\end{abstract}

\maketitle

\section{Introduction}
\label{sec:intro}

The Aharonov-Bohm effect (AB effect), first predicted by Aharonov and Bohm~\cite{Aharonov:1959fk,Chambers:1960xlk}, represents a fundamental quantum phenomenon where charged particles acquire topology-dependent phases when traversing regions with vanishing magnetic field $\mathbf{B}$ but nonvanishing vector potential $\mathbf{A}$. This effect establishes the physical significance of the vector potential in quantum mechanics, demonstrating that particles experience observable phase shifts mediated by $\mathbf{A}$ despite the absence of local electromagnetic forces. The AB effect highlights the role of gauge structures in quantum physics, with precision experiments that confirm both magnetic and electric variants through measurements of scalar and vector potential contributions~\cite{Aharonov:1984xb,vanOudenaarden}.

Recent theoretical developments have explored gravitational analogs of the AB effect. Early foundational work by Ford et al.~\cite{LHFord} employed tetrad-based gravity to formulate covariant Dirac equations (extended to DKP formalisms~\cite{FALEKM}) for spin $1/2$ particles in curved spacetime\cite{Bezerra:1988hx,Boumali:2017xxq}. Subsequent approaches include energy-mass equivalence methods, which derive gravitational AB phases from translation operators~\cite{Tobar_2024}, and Newtonian-potential-based calculations for particles in terrestrial orbits~\cite{PhysRevD.109.064073}. These models establish gravitational counterparts to the electromagnetic AB effect but often rely on specific spacetime geometries or non-relativistic approximations.

Experimental advances, particularly in atom interferometry, now enable precision tests of gravitational phase shifts~\cite{KULIK1996252,Overstreet:2021hea}. By measuring matter-wave interference patterns while suppressing non-gravitational contributions, such experiments probe quantum-gravity interactions at laboratory scales. The phase difference $\Delta\phi$ between interferometer paths is given by~\cite{Overstreet:2021hea}
\begin{equation}
\Delta\phi = \frac{m}{\hbar} \int \left[
V_1(\mathbf{x},t) - V_2(\mathbf{x},t) -
\frac{\Delta x}{2} \left(
\frac{\partial V_1}{\partial x} + \frac{\partial V_2}{\partial x}
\right) \right] dt,
\label{eq:phase_diff}
\end{equation}
where $V_i$ denotes the gravitational potential along each path $i$. This expression captures how potential differences induce measurable quantum phases even without local forces.

Despite these advances, a unified framework for electromagnetic and gravitational AB effects in full four-dimensional spacetime remains elusive. In this work, we study an approach using fiber bundle theory. By treating connections as gauge potentials and computing the holonomy through covariant derivatives, we derive the AB phase for particles following spacetime geodesics. Our formulation establishes gravity as a gauge theory with curvature as field strength, computes holonomy transformations for wavefunctions, and yields gauge-invariant phases valid for both electromagnetic and gravitational interactions. This framework provides useful insights for interpreting current experiments and exploring quantum-gravity phenomena in controlled settings.

This work is structured as follows: section~\ref{sec:em_ab} discusses the geometric formulation of electromagnetic AB phases using $U(1)$ bundle theory, deriving the spacetime decomposition of holonomy. Section~\ref{sec:grav_ab} establishes the gravitational analog through Levi-Civita connections in curved spacetime, demonstrating the Newtonian limit's isomorphism with the electromagnetic case. Section~\ref{sec:summary} summarizes these results, highlighting their implications for quantum-gravity experiments. 

\section{The AB Effect under Electromagnetic Interaction in Four-Dimensional Spacetime}
\label{sec:em_ab}

Within the gauge theory framework, the wave function of a charged particle corresponds to a section of a $U(1)$ principal bundle $\pi : E \to M$, where $E$ denotes the total space and $M$ the four-dimensional spacetime base manifold. Locally, the bundle admits a trivialization such that points are represented as $(p, u_p)$ with $u_p \in \mathbb{C}$. For a charged particle traversing a worldline $\gamma(\tau)$, the parallel transport equation governs the evolution of the wave function,
\begin{equation}
u(0) = u, \quad D_{\gamma}u(\tau) = 0,
\label{eq:parallel_transport}
\end{equation}
where $D$ denotes the covariant derivative of section $s$, defined by $s:M \to E$, which can be written as $(p,u_p)$ locally. According to fiber bundle theory, there is $D_{\gamma}{u_p}=\partial_\tau u_p+A(\tau)u_p$ and $A$ is the connection. In the U(1) bundle theory, it can be proved that connection A is the potential of  electromagnetic interaction. Solving eq.~\eqref{eq:parallel_transport} yields the path-ordered expression
\begin{equation}
u(\tau) = \mathcal{P} \exp\left( \int_{0}^{\tau} A_\mu(\gamma(s)) ds \right) u,
\label{eq:path_ordered}
\end{equation}
where $\mathcal{P}$ indicates path ordering and $A_\mu$ represents the $U(1)$ connection 1-form.

The holonomy group element $H(\gamma, D)$ maps the initial vector $u_p$ at the point $p$ to the final vector $u_q$ at $q$ via $u_q = H(\gamma, D) u_p$. For a closed spacetime loop composed of infinitesimal segments in the $x^\mu$-$x^\nu$ plane, the holonomy is given by
\begin{equation}
H(\gamma, D) = 1 - \epsilon^2 F_{\mu\nu} + \mathcal{O}(\epsilon^3),
\label{eq:infinitesimal_holonomy}
\end{equation}
where $F_{\mu\nu} = \partial_\mu A_\nu - \partial_\nu A_\mu$ is the electromagnetic field tensor.

Consider now the classic AB configuration where a charged particle circumnavigates an infinite solenoid with static uniform magnetic flux. The worldline $\gamma$ connects $(t, \mathbf{r})$ to $(t+T, \mathbf{r})$, and may be decomposed into temporal and spatial components
\begin{equation}
H(\gamma, D) = \prod_{i=1}^n H(\gamma_{C_i}, D) H(\sigma_i, D),
\label{eq:holonomy_decomposition}
\end{equation}
where $\gamma_{C_i}$ denote infinitesimal spatial curves and $\sigma_j$ temporal segments. Under static conditions ($\partial_0 A_i = 0$, $A_0 = \text{const}$), the electric field components vanish, $F_{i0} = 0$. This implies commutation of spatial and temporal holonomies
\begin{equation}
H(\gamma_{C_i}, D) H(\sigma_j, D) = H(\sigma_j, D) H(\gamma_{C_i}, D),
\label{eq:commutation}
\end{equation}
as verified through the infinitesimal loop relation
\begin{equation}
    \begin{aligned}
    H(\gamma_{C_i}, D) &H(\sigma_j, D) H(\gamma_{C_i}^{-1}, D) H(\sigma_j^{-1}, D) \\
    &= 1 - \epsilon^2 F_{ij} \\
    & = H(\gamma_{C_i}, D) H(\gamma_{C_i}^{-1}, D) H(\sigma_j, D) H(\sigma_j^{-1}, D).
\label{eq:loop_identity}
\end{aligned}
\end{equation}

The factorization in eq.~\eqref{eq:commutation} allows the total holonomy to separate as
\begin{equation}
H(\gamma, D) = H(\gamma_C, D) H(\sigma, D),
\label{eq:holonomy_factorization}
\end{equation}
where $\gamma_C = \prod_i \gamma_{C_i}$ is the closed spatial path and $\sigma = \prod_j \sigma_j$ the temporal trajectory. Explicitly, their expressions read
\begin{align}
H(\gamma_C, D) &= \mathcal{P} \exp\left( \oint_{\gamma_C} A_i d\mathbf{r}^i \right),
\label{eq:spatial_holonomy} \\
H(\sigma_C, D) &= \mathcal{P} \exp\left( \int_{\sigma} \Phi(t) dt \right) ,
\label{eq:temporal_holonomy}
\end{align}
with $\Phi(t)$ denoting the electric scalar potential. The final wavefunction at $q$ is then
\begin{equation}
u_q = \mathcal{P} \left[ \exp\left( \oint_{\gamma_C} A_i d\mathbf{r}^i \right)  \exp\left( \int_{\sigma} \Phi(t) dt \right)\right] u_p.
\label{eq:total_phase}
\end{equation}
The first factor in eq.~\eqref{eq:total_phase} corresponds to the magnetic AB phase arising from the vector potential in a field-free region, while the second represents the Electric Aharonov-Bohm (EAB)\cite{Ludwin:2010aa} phase induced by the scalar potential. This demonstrates the unified origin of both effects in the holonomy of the $U(1)$ connection over the spacetime trajectory of a particle.

\section{The Gravitational AB Effect}
\label{sec:grav_ab}

The gravitational analog of the AB effect emerges from the parallel transport of vectors in curved spacetime. For a particle following a worldline $\gamma(\tau)$ with tangent vector $T^b$, the parallel transport equation for a vector $v^a$ is given by
\begin{equation}
T^b \nabla_b u^a = 0, \quad u^a(0) = u^a,
\label{eq:grav_parallel}
\end{equation}
where $\nabla_b$ denotes the covariant derivative associated with the Levi-Civita connection, which has $\nabla_b u^{a}=\partial_{b}u^{a} +\Gamma^{a}_{bc}u^{c}$. The solution to this differential equation can be formally expressed as
\begin{equation}
u^a(t) = \mathcal{P} \exp\left( \int_0^t \Gamma^a_{bc}(\gamma(s)) T^b(s) ds \right) u^c,
\label{eq:grav_solution}
\end{equation}
where $\mathcal{P}$ indicates path ordering and $\Gamma^a_{bc}$ are the Christoffel symbols.

For an infinitesimal loop $\Gamma$ containing point $x$, the holonomy is determined by the Riemann curvature tensor~\cite{Weinberg1972-WEIGAC}
\begin{equation}
H(\Gamma, \nabla) S^\mu = S^\mu - \frac{1}{2} R^{\mu}{}_{\omega\nu\rho} S^\omega \oint dx^\nu x^\rho + \mathcal{O}(\epsilon^3),
\label{eq:infinitesimal_holonomy_grav}
\end{equation}
where $R^{\mu}{}_{\omega\nu\rho}$ is the Riemann tensor and $S^{\mu}$ is a vector field. The parallel transport equation of the latter can be rewritten as 
\begin{equation}
    \frac{dS_{\mu}}{d\tau}=\Gamma^{\lambda}_{\nu \mu}\frac{dx^{\nu}}{d \tau}S_{\lambda}.
\end{equation}

We now specialize to the Newtonian limit~\cite{Frittelli:1994stb}, where the metric perturbation $h_{ab}$ satisfies $|h_{ab}| \ll 1$,
\begin{equation}
g_{ab} = \eta_{ab} + h_{ab}.
\label{eq:linearized_metric}
\end{equation}
Defining the trace-reversed perturbation $\bar{h}_{ab} \equiv h_{ab} - \frac{1}{2} \eta_{ab} h$ and imposing the Lorentz gauge condition $\partial^b \bar{h}_{ab} = 0$, the linearized Einstein equations reduce to
\begin{equation}
\Box \bar{h}_{ab} = -16\pi T_{ab}\, .
\label{eq:linearized_einstein}
\end{equation}
In the Newtonian limit, this yields the Poisson equation for the gravitational potential
\begin{equation}
\nabla^2 \phi = 4\pi \rho, \quad \phi \equiv -\frac{1}{4} \bar{h}_{00},
\label{eq:newtonian_potential}
\end{equation}
where $\rho = T_{00}$ is the mass density.

Consider a particle traversing a path $\gamma$ from $p = (t, \mathbf{x})$ to $q = (t+T, \mathbf{x})$ in an inertial coordinate system $(t, x^i)$, the holonomy decomposes as
\begin{equation}
H(\gamma, \nabla) = \prod_{i=1}^n \left[ H(\gamma_{C_i}, \nabla) H(\sigma_i, \nabla) \right],
\label{eq:holonomy_decomposition_grav}
\end{equation}
where $\gamma_{C_i}$ and $\sigma_i$ are spatial and temporal segments, respectively. Under Newtonian conditions, the curvature components $R_{ab0i}$ vanish, implying the commutation of spatial and temporal holonomies,
\begin{equation}
H(\gamma_{C_i}, \nabla) H(\sigma_j, \nabla) = H(\sigma_j, \nabla) H(\gamma_{C_i}, \nabla).
\label{eq:commutation_grav}
\end{equation}
The total holonomy then factorizes as
\begin{equation}
H(\gamma, \nabla) = H(\gamma_C, \nabla) H(\sigma, \nabla),
\label{eq:holonomy_factorization_grav}
\end{equation}
where $\gamma_C = \prod_i \gamma_{C_i}$ is a closed spatial path and $\sigma = \prod_j \sigma_j$ is the temporal path.

The spatial holonomy $H(\gamma_C, \nabla)$ corresponds to parallel transport around a closed spatial loop. Applying Stokes' theorem to the curvature 2-form
\begin{equation}
H(\gamma_C, \nabla) = \mathcal{P} \exp\left( -\frac{1}{2} \int_U R^{\mu}{}_{\nu\rho\sigma} dS^{\rho\sigma} \right), \quad 
\label{eq:spatial_holonomy_grav}
\end{equation}
where $U$ is a 2-surface bounded by $\gamma_C$, and $dS^{\rho\sigma}$ is the surface element. This phase shift represents the gravitational analog of the magnetic AB effect.

For the temporal holonomy, the phase shift is expressed by
\begin{equation}
    H(\sigma,\nabla)= \prod H(\sigma_{i},\nabla)= \prod \left[\Gamma^{a}_{0c}(x)(t_{i+1}-t_{i})\right].
\end{equation}
This gives
\begin{equation}
    H(\sigma,\nabla) = \exp\left(\int^{T}_{0} \ln \left[1-\left(1-\Gamma^{a}_{0c}\left(x(t) \right) \right)\right]dt \right) .
    \label{eq:Hsigmanabla}
\end{equation}
Notice that the exponential part of the first term, which contributes to the phase shift, corresponds to the space curve $\int R^{0}{}_{\mu \nu \rho }dS^{\nu \rho }$. Furthermore, as a remark, the dominant Christoffel symbol in the Newtonian limit is small enough, which gives the following:
\begin{equation}
H(\sigma, \nabla) =  \exp\left[ \int_0^{T}\left( \Gamma^a_{0c}-1\right) dt \right] .
\label{eq:temporal_holonomy_grav}
\end{equation}
This phase shift corresponds to the gravitational analog of the electric AB effect. The nonzero components of the Christoffel symbol under the Newtonian limit are
$\Gamma^{i}_{0i}=-\left(1+2\phi\right)\partial_{0}\phi$, $\Gamma^{i}_{00}=\left(1+2\phi\right)\partial_{i}\phi$, $\Gamma^{i}_{00}=\left(2\phi-1\right)\partial_{0}\phi$, and $\Gamma^{0}_{0i}=\left(1-2\phi\right)\partial_{i}\phi$.

In general, the part under the integration in eq.~\eqref{eq:Hsigmanabla} can be viewed as the components of an 02 tensor $\omega_{ab}$. And this tensor is antisymmetric about swapping its two indices. With this property, the phase shift exponential term can be rewritten as

\begin{equation}
    H(\sigma, \nabla) =  \frac{1}{2}\int _{U}\omega_{ab},
\end{equation}
where $\omega_{ab}$ is a 2-form and $ \gamma _{C} $ can be described as a 2-dimensional manifold, on which the integration is defined. Due to the Stoke's theorem, we have
\begin{equation}
    \frac{1}{2}\int_{U}\omega_{ab}=\frac{1}{2}\int_{\partial U}S_{a}, 
\end{equation}
where $dS_{c}=\omega_{ab}$ and $S_{a} \in \Omega^{1}$. Here, we note that $\partial U=\gamma_{C}$,  the final expression for the phase shift exponential term becomes 
\begin{equation}
    H(\sigma, \nabla) = \frac{1}{2}\int_{\gamma_{C}}S_{a}.
\end{equation}

Combining these, we obtain 
\begin{equation} 
H(\gamma,\nabla) =\mathcal{P} \exp\left[ \int_0^{T}\left( \Gamma^a_{0c}-1\right) dt \right]\exp\left( \frac{1}{2}\int_{\gamma_{C}}S_{a}\right).
\label{eq:gravab_total}
\end{equation} 

The total phase shift for the wavefunction is:
\begin{equation}
u_q = \mathcal{P} \exp\left[ \int_0^{T}\left( \Gamma^a_{0c}-1\right) dt \right] \exp\left( \frac{1}{2}\int_{\gamma_{C}}S_{a}\right) u_p.
\label{eq:total_grav_phase}
\end{equation}

The first term represents curvature-induced phase shifts in vacuum regions (gravito-magnetic AB effect), while the second arises from gravitational potential gradients (gravito-electric AB effect). This demonstrates the unified geometric origin of gravitational AB phenomena in four-dimensional spacetime.

\section{Summary}
\label{sec:summary}

In this work, we have studied a unified geometric framework for analyzing AB effects in a four-dimensional spacetime. Beginning with the electromagnetic case, we demonstrated that for a charged particle moving around a solenoid, the phase shift factorizes into distinct spatial and temporal components, as shown in eq.~\eqref{eq:total_phase}. The first term corresponds to the magnetic AB phase arising from the vector potential, while the second represents the electric EAB phase induced by the scalar potential.

We subsequently extended this analysis to gravitational interactions, deriving the explicit phase shift under Newtonian limit conditions. The gravitational holonomy similarly decomposes into spatial and temporal contributions as expressed in eq.~\eqref{eq:total_grav_phase}. The temporal term involves the integral of the 1-form $S_a$ along the closed curve $\gamma_{C}$, while the spatial term depends on a 2-form in a two dimensional manifold. The latter depends on the Christoffel connection integrated over the worldline. This factorization mirrors the electromagnetic case, revealing a structural parallel between the two interactions.

The isomorphism between eqs.~\eqref{eq:total_phase} and \eqref{eq:total_grav_phase} demonstrates that both electromagnetic and gravitational AB effects originate from the holonomy of their respective connections: the $U(1)$ gauge potential in electromagnetism and the Levi-Civita connection in gravity. In each case, the total phase shift separates into curvature-dependent (spatial) and potential-dependent (temporal) components when static field conditions are satisfied. This geometric unification provides a  foundation for interpreting quantum interference phenomena in curved spacetime and establishes a theoretical framework for ongoing experimental investigations of gravito-AB effects using atom interferometry techniques.

\textbf{Acknowledgment:} This work was supported by the Natural Science Foundation of Xinjiang Uygur Autonomous Region of China under Grant No.~2022D01C52, and by the "Quantum Field Theory Excellent Course" Project of Xinjiang University, under Grant No.~XJDX2023YJPK07.
%

\bibliography{apssamp}

\end{document}